\documentclass[final]{aipproc}

\layoutstyle{8x11single}
\usepackage{amsfonts}
\usepackage{amssymb}
\def\ka{\left.\mid 11 \right\rangle_c}
\def\kb{\left.\mid 88 \right\rangle_c}
\def\kap{\left.\mid 1'1' \right\rangle_c}
\def\kbp{\left.\mid 8'8' \right\rangle_c}
\def\ba{\,_c\left\langle 11 \mid\right.}
\def\bb{\,_c\left\langle 88 \mid\right.}
\def\bap{\,_c\left\langle 1'1' \mid\right.}
\def\bbp{\,_c\left\langle 8'8' \mid\right.}
\def\paap{\,_c\left\langle 11 \mid 1'1' \right\rangle_c}

\begin{document}

\title{Meson-Meson molecules and compact four-quark states}


\author{J. Vijande}{
  address={Departamento de F\'{\i}sica At\'{o}mica, Molecular y Nuclear, Universidad de Valencia (UV)  
and IFIC (UV-CSIC), Valencia, Spain.}}

\author{A. Valcarce}{
  address={Departamento de F\'{\i}sica Fundamental, Universidad de Salamanca, Salamanca, Spain.}}

\begin{abstract}
The physics of charm has become one of the best laboratories exposing the
limitations of the naive constituent quark model and also giving hints
into a more mature description of meson spectroscopy, beyond the simple quark--antiquark configurations. In this talk we review some recent studies of multiquark components in 
the charm sector and discuss in particular
exotic and non-exotic four-quark systems.
\end{abstract}

\maketitle

More than thirty years after the so-called November revolution~\cite{Bjo85}, 
heavy hadron spectroscopy remains a challenge. The formerly comfortable 
world of heavy mesons is shaken by new 
results~\cite{Ros07}. This  started in 2003 
 with the discovery of the $D_{s0}^*(2317)$ and  $D_{s1}(2460)$ 
 mesons in the open-charm sector. These positive-parity states have masses lighter than expected from quark  models, and
also smaller widths. Out of the many proposed explanations, 
the unquenching of the naive quark model has been successful~\cite{Vij05}. 
When a $(q\bar q)$ pair occurs in a $P$-wave but can couple to hadron 
pairs in $S$-wave, the latter configuration distorts the $(q\bar q)$ picture. 
Therefore, the $0^+$ and $1^+$ $(c\bar s)$ states
predicted above the $DK(D^*K)$ thresholds couple to the continuum. This 
mixes meson--meson components in the wave function,  an idea advocated long 
ago to explain the spectrum and properties 
of light-scalar mesons~\cite{Jaf07}.

This possibility of $(c\bar{s} n\bar{n})$ ($n$ stands for a light quark) components in $D_s^*$ has open the discussion about the presence of compact 
$(c\bar c n\bar n)$ 
four-quark states in the charmonium spectroscopy.  Some states recently found in the hidden-charm sector may fit in the 
simple quark-model description as $(c\bar c)$  pairs 
(e.g., $X(3940)$, $Y(3940)$, and $Z(3940)$ 
as radially excited  $\chi_{c0}$, $\chi_{c1}$, and $\chi_{c2}$), but
 others appear to be more elusive, in particular $X(3872)$, $Z(4430)^+$, and $Y(4260)$. 
The  debate on the nature of these states is open,   with special emphasis on the 
$X(3872)$.
Since it was first reported by Belle in 2003~\cite{Bel03}, 
it has gradually become the flagship of the new armada of states 
whose properties make their identification as
traditional $(c\bar c)$ states unlikely.
An average mass of $3871.2\pm0.5\;$MeV and a narrow width of less than $2.3\;$MeV
have been reported for the $X(3872)$.
Note the vicinity of this state to the $D^0\overline{D}{}^{*0}$ threshold,
$M(D^0\,\overline{D}{}^{*0})=3871.2\pm1.2\;$MeV.
With respect to the $X(3872)$ quantum numbers, although some caution is still 
required until better statistic
is obtained~\cite{Set06}, an isoscalar $J^{PC}=1^{++}$ state seems to
be the best candidate to describe the properties of the $X(3872)$.

Another hot sector, at least for theorists, includes the $(cc\bar n\bar n)$ states, which are manifestly exotic with 
charm $2$ and  baryon number $0$. Should they lie below the threshold for dissociation into two ordinary hadrons, they would 
be narrow and show up clearly in the experimental 
spectrum. There are already estimates of the production rates
indicating they could be produced and detected at present 
(and future) experimental facilities~\cite{Fab06}.
The stability of such $(QQ\bar q\bar q)$ states has been discussed since the early 80s \cite{Ade82}, and there is a consensus that 
stability is reached when the mass ratio $M(Q)/m(q)$ becomes large enough.  See, e.g., \cite{Janc} for Refs. This effect is also 
found in QCD sum rules \cite{Navarra:2007yw}.
This  improved binding when $M/m$ increases is due to the same mechanism by which the hydrogen molecule $(p,p,e^-,e^-)$ is much 
more bound than the positronium molecule $(e^+,e^+,e^-,e^-)$. What matters is not the Coulomb character of the potential, but its 
property to remain identical when the masses change. In quark physics, this property is named \emph{flavour independence}.  It is 
reasonably well satisfied, with departures mainly due to spin-dependent corrections.

The question is whether stability is already possible for $(cc\bar n\bar n)$ or requires heavier quarks. 
In Ref.~\cite{Janc}, a marginal binding was found for a specific potential for which earlier studies found no binding. This 
illustrates how difficult are such four-body calculations.

Besides trying to unravel the possible existence of
bound $(cc\bar n\bar n)$ and $(c\bar c n\bar n)$ states
one should aspire to understand whether it is possible to differentiate between compact
and molecular states. A molecular state may be understood as a four-quark state
containing a single physical two-meson component, i.e., a unique singlet-singlet
component in the colour wave function with well-defined spin and isospin quantum numbers.
One could expect these states not being deeply bound and therefore having a size
of the order of the two-meson system, i.e., 
$\Delta_R\sim1$. Opposite to that,
a compact state may be characterized by its involved structure on the
colour space, its wave function containing different singlet-singlet
components with non negligible probabilities. One would expect
such states would be smaller than typical two-meson systems, i.e.,
$\Delta_R < 1$. Let us notice that while $\Delta_R>1$ but finite would 
correspond to a meson-meson molecule 
$\Delta_R\stackrel{K\to\infty}{\longrightarrow}\infty$ 
would represent an unbound threshold. Thus, dealing with four-quark
states an important question is whether we are in front of a
colorless meson-meson molecule or a compact state, i.e., a system
with two-body colored components. While the first structure would be
natural in the naive quark model, the second one would open a new
area on the hadron spectroscopy.

There are three different ways of coupling two quarks and two antiquarks 
into a colorless state:

\begin{eqnarray}
\label{eq1a}
[(q_1q_2)(\bar q_3\bar q_4)]&\equiv&\{|\bar 3_{12}3_{34}\rangle,|6_{12}\bar 6_{34}\rangle\}\equiv\{|\bar 33\rangle_c^{12},
|6\bar 6\rangle_c^{12}\}\\
\label{eq1b}
[(q_1\bar q_3)(q_2\bar q_4)]&\equiv&\{|1_{13}1_{24}\rangle,|8_{13} 8_{24}\rangle\}\equiv\{|11\rangle_c,|88\rangle_c\}\\
\label{eq1c}
[(q_1\bar q_4)(q_2\bar q_3)]&\equiv&\{|1_{14}1_{23}\rangle,|8_{14} 8_{23}\rangle\}\equiv\{|1'1'\rangle_c,|8'8'\rangle_c\}\,,
\label{eq1}
\end{eqnarray}
\noindent
being the three of them orthonormal basis. Each coupling scheme allows to 
define a color basis where the four--quark problem can be solved. 
The first basis, Eq.~(\ref{eq1a}), 
being the most suitable one to deal with the Pauli principle is made 
entirely of vectors containing hidden--color components. The other two, Eqs.~(\ref{eq1b}) and~(\ref{eq1c}), 
are hybrid basis containing singlet--singlet (physical) and octet--octet (hidden--color) 
vectors.

To evaluate the probability of physical channels (singlet-singlet
color states) one needs to expand any hidden-color vector of the
four-quark state color basis in terms of singlet-singlet color
vectors. Given a general four-quark state this requires to mix terms
from two different couplings, \ref{eq1b} and~\ref{eq1c}. 
In ~\cite{Vij09c} the two Hermitian operators that are well-defined
projectors on the two physical singlet-singlet color states were
derived,
\begin{eqnarray}
{\cal P}_{\ka} & =&  \left( P\hat Q + \hat Q P \right) \frac{1}{2(1-|\paap|^2)}
\nonumber \\
{\cal P}_{\kap} & =&  \left( \hat P Q + Q \hat P \right) \frac{1}{2(1-|\paap|^2)}  \, ,
\label{tt}
\end{eqnarray}
where $P$, $Q$, $\hat P$, and $\hat Q$ are the projectors over the
basis vectors (\ref{eq1b}) and (\ref{eq1c}),
\begin{eqnarray}
P & = & \ka \ba \nonumber \\
Q & = & \kb\bb  \, ,
\label{Proj1}
\end{eqnarray}
and
\begin{eqnarray}
\hat P & = & \kap\bap \nonumber \\
\hat Q & = & \kbp\bbp  \, .
\label{Proj2}
\end{eqnarray}

By using them and the formalism of ~\cite{Vij09c}, the four-quark nature (unbound, molecular
or compact) can be explored. Such a formalism can be applied to any four-quark
state, however, it becomes much simpler when distinguishable quarks are present. This would be,
for example, the case of the $(nQ\bar n\bar Q)$ system, where the Pauli principle does not apply.
In this system the bases (\ref{eq1b}) and (\ref{eq1c}) are distinguishable due to the flavor part,
they correspond to $[(n\bar c)(c\bar n)]$ and $[(n\bar n)(c\bar c)]$,
and therefore they are orthogonal. This makes that the probability of a physical channel
can be evaluated in the usual way for
orthogonal basis~\cite{Vij07}. The non-orthogonal bases formalism is required for those cases
where the Pauli Principle applies either for the quarks or the antiquarks pairs.
Relevant expressions can be found in~\cite{Vij09c}. We
show in Table~\ref{re1} some examples of results obtained for
heavy-light tetraquarks. One can see how independently of their
binding energy, all of them present a sizable octet-octet component
when the wave function is expressed in the~\ref{eq1b} coupling. Let
us first of all concentrate on the two unbound states, $\Delta_E >
0$, one with $S=0$ and one with $S=1$, given in Table~\ref{re1}. The
octet-octet component of basis~(\ref{eq1b}) can be expanded in terms
of the vectors of basis~(\ref{eq1c}) as explained in the previous
section. Then, the probabilities are concentrated into a single
physical channel, $MM$ or $MM^*$ [$MM$ stands for two identical
pseudoscalar $D$ ($B$) mesons and $MM^*$ for a pseudoscalar $D$
($B$) meson together with its corresponding vector excitation, $D^*$
($B^*$)]. In other words, the octet-octet component of the
basis~(\ref{eq1b}) or~(\ref{eq1c}) is a consequence of having
identical quarks and antiquarks. Thus, four-quark unbound states are
represented by two isolated mesons. This conclusion is strengthened
when studying the root mean square radii, leading to a picture where
the two quarks and the two antiquarks are far away, $\langle
x^2\rangle^{1/2}\gg 1$ fm and $\langle y^2\rangle^{1/2}\gg 1$ fm,
whereas the quark-antiquark pairs are located at a typical distance
for a meson, $\langle z^2\rangle^{1/2}\le 1$ fm. Let us now turn to
the bound states shown in Table~\ref{re1}, $\Delta_E < 0$, one in
the charm sector and two in the bottom one. In contrast to the
results obtained for unbound states, when the octet-octet component
of basis~(\ref{eq1b}) is expanded in terms of the vectors of
basis~(\ref{eq1c}), one obtains a picture where the probabilities in
all allowed physical channels are relevant. It is clear that the
bound state must be generated by an interaction that it is not
present in the asymptotic channel, sequestering probability from a
single singlet-singlet color vector from the interaction between
color octets. Such systems are clear examples of compact four-quark
states, in other words, they cannot be expressed in terms of a
single physical channel.

\begin{figure}
{\includegraphics[width=.5\textwidth]{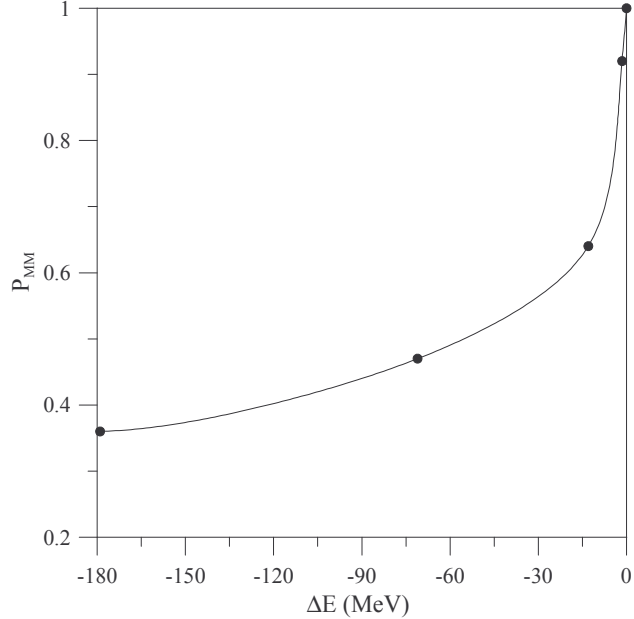}}
\caption{\label{f2}$P_{MM}$ as a function of $\Delta_E$.}
\label{fignew} 
\end{figure}

We have studied the dependence of the probability of a physical channel
on the binding energy. For this purpose we have considered the simplest
system from the numerical point of view, the
$(S,I)=(0,1)$ $cc\bar n\bar n$ state. Unfortunately, this state
is unbound for any reasonable set of parameters. Therefore, we bind it by multiplying the
interaction between the light quarks by a fudge factor.
Such a modification does not affect the two-meson threshold while
it decreases the mass of the four-quark state. The results are illustrated in
Figure~\ref{fignew}, showing how in the $\Delta_E\to0$ limit,
the four-quark wave function is almost a pure single physical
channel. Close to this limit one would find what could be defined as
molecular states. When the probability concentrates
into a single physical channel ($P_{M_1M_2}\to 1$) the
system gets larger than two isolated mesons~\cite{Vij09c}.
One can identify the subsystems responsible for increasing the size of the four-quark state.
Quark-quark ($\langle x^2\rangle^{1/2}$) and antiquark-antiquark ($\langle y^2\rangle^{1/2}$)
distances grow rapidly while the quark-antiquark  distance ($\langle z^2\rangle^{1/2}$)
remains almost constant. This reinforces our previous result, pointing to the appearance
of two-meson-like structures whenever the binding energy goes to zero.

\begin{table}
\caption{Heavy-light four-quark state properties for selected
quantum numbers. All states have positive parity and total orbital
angular momentum $L=0$. Energies are given in MeV. The notation
$M_1M_2\mid_{\ell}$ stands for mesons $M_1$ and $M_2$ with a
relative orbital angular momentum $\ell$. $P[| \bar 3
3\rangle_c^{12}(| 6\bar 6\rangle_c^{12})]$ stands for the
probability of the $3\bar 3(\bar 6 6)$ components given in
Equation~(\ref{eq1a}) and $P[\ka(\kb)]$ for the $11(88)$ components
given in Equation~(\ref{eq1b}). $P_{MM}$, $P_{MM^*}$, and
$P_{M^*M^*}$ have been calculated following the formalism
of~\cite{Vij09c}, and they represent the probability of finding
two-pseudoscalar ($P_{MM}$), a pseudoscalar and a vector
($P_{MM^*}$) or two vector ($P_{M^*M^*}$) mesons}. 
\label{re1}
\begin{tabular}{|c|ccccc|}
\hline
$(S,I)$             & (0,1)      &  (1,1)      & (1,0)     & (1,0)      & (0,0) \\
Flavor              &$cc\bar n\bar n$&$cc\bar n\bar n$&$cc\bar n\bar n$&$bb\bar n\bar n$&$bb\bar n\bar n$\\
\hline
Energy              & 3877       &  3952       & 3861      & 10395      & 10948 \\
Threshold           & $DD\mid_S$     &  $DD^*\mid_S$   & $DD^*\mid_S$   & $BB^*\mid_S$   &  $B_1B\mid_P$\\
$\Delta_E$              & +5         &  +15        & $-76$     & $-$217     &  $-153$ \\
\hline
$P[| \bar 3 3\rangle_c^{12}]$   & 0.333      &  0.333      & 0.881     & 0.974      &  0.981 \\
$P[| 6 \bar 6\rangle_c^{12}]$   & 0.667      &  0.667      & 0.119     & 0.026      &  0.019 \\
\hline
$P[\ka]$            & 0.556      &  0.556      & 0.374     & 0.342      &  0.340 \\
$P[\kb]$            & 0.444      &  0.444      & 0.626     & 0.658      &  0.660 \\
\hline
$P_{MM}$            & 1.000      &  $-$        & $-$       & $-$        &  0.254 \\
$P_{MM^*}$          & $-$        &  1.000      & 0.505     & 0.531      &  $-$ \\
$P_{M^*M^*}$            & 0.000      &  0.000      & 0.495     & 0.469      &  0.746 \\
\hline
\end{tabular}
\end{table}

In another recent investigation, the four-body  Schr\"odinger equation
has been solved accurately using the hyperspherical harmonic (HH) formalism~\cite{Vij07}, with 
 two standard quark models
containing a linear confinement supplemented by a Fermi--Breit one-gluon exchange
interaction (BCN), and also boson exchanges
between the light quarks (CQC). The model parameters were tuned
in the meson and baryon spectra. 
The results are given in Table~\ref{t1}, indicating the quantum numbers of the
state studied, the maximum value of the grand angular momentum used in the HH 
expansion, $K_{\rm max}$, and the
energy difference between the mass of the 
four-quark state, $E_{4q}$, and that of the lowest two-meson
threshold calculated with the same potential model, $\Delta_E$. For
the $(cc\bar n \bar n)$ system we have also calculated 
the radius of the four-quark
state, $R_{4q}$, and its ratio to the sum of the radii of the 
lowest two-meson threshold, $\Delta_R$.

\begin{table}[t]
\centering
\caption{ $(c\bar c n\bar n$) and $(c c \bar n\bar n$) results.}
\begin{tabular}{|c|cc|cc||c|c|cccc|} 
\hline
\multicolumn{5}{|c||}{$(c\bar c n\bar n$)} &\multicolumn{6}{|c|}{$(c c \bar n\bar n$)} \\
\hline
 &\multicolumn{2}{|c|}{CQC} &\multicolumn{2}{|c||}{BCN}  & & &\multicolumn{4}{|c|}{CQC} \\
\hline
$J^{PC}(K_{\rm max})$ & $E_{4q}$ & $\Delta_{E}$&$E_{4q}$ & $\Delta_{E}$ &
& $J^{P}(K_{\rm max})$ & $E_{4q}$ & $\Delta_{E}$ &$R_{4q}$ & $\Delta_R$ \\
\hline
$0^{++}$ (24) & 3779 &  +34 &  3249 &  +75  &
 & $0^{+}$ (28) & 4441 &  +15 &  0.624 & $> 1$ \\
$0^{+-}$ (22) & 4224 &  +64 &  3778 & +140  &
& $1^{+}$ (24) & 3861 &$-$76 &  0.367 & 0.808 \\
$1^{++}$ (20) & 3786 &  +41 &  3808 & +153  &
$I=0$ & $2^{+}$ (30) & 4526 &  +27 &  0.987 & $> 1$ \\
$1^{+-}$ (22) & 3728 &  +45 &  3319 &  +86  &
& $0^{-}$ (21) & 3996 &  +59 &  0.739 & $> 1$ \\
$2^{++}$ (26) & 3774 &  +29 &  3897 &  +23  &
& $1^{-}$ (21) & 3938 &  +66 &  0.726 & $> 1$ \\
$2^{+-}$ (28) & 4214 &  +54 &  4328 &  +32  &
& $2^{-}$ (21) & 4052 &  +50 &  0.817 & $> 1$ \\
\cline{6-11}
$1^{-+}$ (19) & 3829 &  +84 &  3331 & +157  &
& $0^{+}$ (28) & 3905 &  +50 &  0.817 & $> 1$ \\
$1^{--}$ (19) & 3969 &  +97 &  3732 &  +94  &
& $1^{+}$ (24) & 3972 &  +33 &  0.752 & $> 1$  \\
$0^{-+}$ (17) & 3839 &  +94 &  3760 & +105  &
$I=1$ & $2^{+}$ (30) & 4025 &  +22 &  0.879 & $> 1$ \\
$0^{--}$ (17) & 3791 & +108 &  3405 & +172  &
& $0^{-}$ (21) & 4004 &  +67 &  0.814 & $> 1$ \\
$2^{-+}$ (21) & 3820 &  +75 &  3929 &  +55  &
& $1^{-}$ (21) & 4427 &  +1  &  0.516 & 0.876 \\
$2^{--}$ (21) & 4054 &  +52 &  4092 &  +52 &
& $2^{-}$ (21) & 4461 &$-$38 &  0.465 & 0.766 \\
\hline
\end{tabular}
\label{t1}
\end{table}

As can be seen in Table~\ref{t1} (left), in the case of the $(c\bar c n\bar n)$
there appear no compact bound states for any set of quantum numbers, including
the suggested assignment for the $X(3872)$. 
Independently of the quark--quark interaction and the quantum numbers 
considered, the system evolves to a
well separated two-meson state. This is clearly seen
in the energy, approaching the threshold made of two free mesons, 
and also in the probabilities of the
different colour components of the wave function
and in the radius~\cite{Vij07}. Thus, in any manner one can claim for the existence
of a compact bound state for the $(c\bar c n \bar n)$ system. 

A completely different behaviour is observed in Table~\ref{t1} (right).
Here, there are some particular quantum numbers 
where the energy is quickly stabilized below
the theoretical threshold.
Of particular interest is the $1^+$ $cc\bar n\bar n$ state, whose existence 
was predicted more than twenty years ago~\cite{Zou86}. There is a remarkable agreement on the 
existence of an isoscalar $J^P=1^+$ $cc\bar n\bar n$ bound state using both BCN and CQC models, if not in its properties. 
For the CQC model the predicted binding energy is large, $-$ 76 MeV, $\Delta_R<1$, and a very involved structure of its wave function (the $DD^*$ component of its wave function only accounts for the 50\% of the total probability) what
would fit into compact state. Opposite to that, the BCN model
predicts a rather small binding, $-$7 MeV, and $\Delta_R$ is larger than 1, although finite. This state
would naturally correspond to a meson-meson molecule.

Concerning the other two states that are below threshold in Table~\ref{t1} (right) a more careful analysis is required.
Two-meson thresholds must be determined assuming quantum number conservation 
within exactly the same scheme used in the four--quark calculation. 
Dealing with strongly interacting particles, the two-meson states should have well defined total angular 
momentum, parity, and a properly symmetrized wave function if two identical mesons
are considered (coupled scheme). 
When noncentral forces are not taken into account, orbital angular 
momentum and total spin are also good quantum numbers (uncoupled scheme). 
We would like to emphasize that although we use central forces in our calculation the coupled scheme is the relevant one
for observations, since a small non-central component in the
potential is enough to produce a sizeable effect on the width of a state. These state are below the thresholds given by the
uncoupled scheme but above the ones given within the coupled scheme 
what discard these quantum numbers as promising candidates for being observed
experimentally. 

Binding increases for larger $M/m$, but in the $(bb\bar n\bar n)$ sector, there is no proliferation of bound states.
We have studied all ground states of $(bb\bar n\bar n)$ 
using the same interacting potentials as in the double-charm case. Only four bound states have been found,  with quantum numbers $J^P(I)=1^+(0)$
, $0^+(0)$, $3^-(1)$, and $1^-(0)$. The first three ones correspond to compact states.

To conclude, let us stress again the important difference between the two physical systems which have been considered.
While for the $(c\bar c n\bar n)$, there are two allowed physical {\it decay channels},
$(c\bar c)+(n\bar n)$ and $(c\bar n)+(\bar c n)$, for the $(cc\bar n\bar n)$
only one physical system contains 
the possible final states, $(c \bar n)+(c\bar n)$.
Therefore, a $(c \bar c n \bar n)$ four-quark state will hardly present 
bound states, because the system will reorder itself to become the lightest two-meson state, either
$(c\bar c)+(n\bar n)$ or $(c\bar n)+(\bar c n)$. In other words,
if the attraction is provided by the interaction between
particles $i$ and $j$, it does also contribute to the asymptotic
two-meson state. This does not happen
for the $(c c\bar n\bar n)$ if the interaction between, for example,
the two quarks is strongly attractive. 
In this case there is no asymptotic two-meson
state including such attraction, and therefore the system might bind.

Once all possible $(cc\bar n\bar n)$, $(bb\bar n\bar n)$ and $(c\bar cn\bar n)$ quantum numbers have 
been exhausted very few alternatives remain. If additional bound four-quark states or higher configuration are  
experimentally found, then other mechanisms should be at work, for instance based on diquarks \cite{Jaf07,Mai04,Ale07}.

This work has been partially funded by the Spanish Ministerio de
Educaci\'on y Ciencia and EU FEDER under Contract No. FPA2007-65748,
by Junta de Castilla y Le\'{o}n under Contract No. SA016A17, and by the
Spanish Consolider-Ingenio 2010 Program CPAN (CSD2007-00042).

\bibliographystyle{aipproc}   

\end{document}